\documentclass[3p,twocolumn,preprint]{elsarticle}
\usepackage{ecrc}
\volume{00}
\firstpage{1}
\journalname{Optics Communications}
\runauth{M Mehta, J T Andrews, P Sen}
\jid{oc}
\jnltitlelogo{\small Optics Communications}
\CopyrightLine{2024}{Published by Elsevier Ltd.}
\usepackage{graphicx}
\usepackage{amsmath,amssymb}
\begin{document}
\begin{frontmatter}
\title{Quantum Entanglement in  a Diluted Magnetic Semiconductor Quantum Dot}
\author[sop]{Manish Kumar Mehta }
\author[sgs]{Joseph Thomas Andrews}
\ead{Corresponding Author: jtandrews@sgsits.ac.in}
\author[sop]{Pratima Sen}       
\address[sop]{Laser Bhawan, School of Physics, Devi Ahilya University, Khandwa Road, Indore, MP, 452007 India.}
\address[sgs]{Dept. Applied Physics \& Optoelectronics, Shri G. S. Institute of Technology \& Science, Indore, MP, 452003 India.}
\begin{abstract} 
We investigated the entanglement in a diluted magnetic semiconductor quantum dot, crucial for quantum technologies. Despite their potential, these systems exhibit low extraction rates. We explore self-assembled InGaAs quantum dots, focusing on entanglement between them based on spin states. Our analysis involves defining wavefunctions, employing density matrix operators, and measuring entanglement entropy. Numerical assessments reveal few promising pairs among various quantum dot combinations that exhibit significant entanglement. Additionally, this work discusses theoretical developments and statistical evaluations of entanglement in diluted magnetic semiconductor quantum dots, providing insights into their potential for quantum applications.
\end{abstract}
\end{frontmatter}
\section{Introduction}
In quantum mechanics, the term ``entanglement'' refers to a strong correlation between the characteristics of two or more particles irrespective of their separation. Particles like electrons, photons, or atoms can exist in a superposition of states, where they are simultaneously in multiple states until they are measured or observed. The behavior of one of the entangled particles is immediately coupled to the behavior of the others, regardless of the distance between them. As a result, when the characteristics of an entangled particle are measured, the characteristics of all the other particle become correlated with that measurement and instantly change to maintain the overall state of entanglement.  This correlation holds true even if the particles are separated in space, suggesting that information is somehow transmitted faster than the speed of light, although this does not violate the principle of causality or allow for faster-than-light communication \cite{einstein1935can}.

Quantum entanglement has been a challenge to our understanding of the fundamental nature of reality ever since it was first proposed by Einstein, Podolsky, and Rosen in their seminal article \cite{einstein1935can}.

The idea of quantum entanglement is  important theoretically, and has undergone rigorous experimental studies. In quantum information science which is the basis of quantum computer, secure communication, quantum teleportation and quantum cryptography, the use of entangled state is of fundamental importance \cite{bennett1984quantum, lewis2019dynamics}. Quantum teleportation, a ground-breaking mechanism that uses the entanglement shared by particles to enable the transfer of quantum states, also depends on quantum entanglement \cite{hu2023progress, bouwmeester2000physics}.  In fact, it is anticipated that quantum entanglement will significantly improve computer performance, secure data network architecture, state teleportation, etc \cite{benjamin2009prospects, bogdanov2011quantum, nielsen2010quantum}. Making quantum gates and contributing significantly to quantum communication and information processing is the concept of quantum entanglement \cite{hughes2021entanglement, hu2021long}. Numerous experiments, including the well-known Bell's theorem experiments that demonstrated the incompatibility of certain local hidden variable theories with the predictions of quantum physics, have been used to scientifically validate the concept of entanglement. For instance, the polarization states of photons and the spin states of electrons. The quantum states of an entangled system cannot be factored as a product of states of its local constituents. Bell established the maximum strength of correlations and demonstrated that certain entangled systems will violate this maximum according to quantum theory \cite{bell1964einstein}. If two photons are entangled, no matter how far apart two photons are from one another, measuring the polarization of one of the photons will reveal the polarization of the other. Kocher et al \cite{kocher1967polarization} showed that calcium atoms can emit two entangled photons. Togan et al \cite{togan2010nature} validates quantum entanglement between optical photon polarization and a solid-state qubit with an electronic spin of a nitrogen vacancy centre in diamond. The quantum eraser technique is used to demonstrate a high degree of control over interactions between a solid state qubit and the quantum light field. Warren et al. \cite{warren2021robust} demonstrated a  photon-mediated cross-resonance gate, which is suitable for realistic experimental capabilities and does not require resonant tuning. Chan et al. \cite{chan2023chip} showcase high-fidelity on-chip entanglement between a photon and a quantum-dot hole spin qubit. This was achieved through rapid photon scattering and active spin control in a microsecond, much quicker than other solid-state platforms.
The interactions that lead to the entangled systems can take many different forms. When there is no interaction between the components of the composite system, quantum entanglement is feasible.

In a bipartite system, quantum entanglement pertains to a system containing two particles or subsystems which could include atoms, photons, or larger objects. Describing the entangled state of a bipartite system requires a joint description which accounts for their entanglement and cannot be described by considering the states of its individual components separately.
A tensor product is commonly used to describe bipartite entangled states in mathematics. An entangled state of two particles, A and B, can be represented in this way.
\begin{equation}\label{eq:1}
|{\psi}\rangle=\alpha|0\rangle_A|1\rangle_B+\beta|1\rangle_A|1\rangle_B,
\end{equation}
here, the states of each particle are represented by $|0\rangle$ and $|1\rangle$, while the different configurations' probability amplitudes are determined by the complex numbers $\alpha$ and $\beta$. The entangled states have crucial characteristics, that individual states of the particles A and B are not well-defined but rather exists in a superposition of various possibilities.

Generally speaking, an entangled state $|{\psi}\rangle$ of a bipartite system can be identified if  condition such as 
\begin{equation}\label{eq:2}
|{\psi}\rangle_{AB}=\sum_{i,j}c_{i,j}|\psi_i\rangle_A\otimes|\psi_j\rangle_B \neq |\psi\rangle_A\otimes|\psi\rangle_B,
\end{equation}
is met \cite{meystre2007elements}. 
Gupta et al. \cite{gupta2015entanglement} conducted a study that delves into the quantification of entanglement in pure bipartite states. This was achieved using the Schmidt number and Schmidt rank \cite{gupta2015entanglement, roncaglia2014bipartite}, which served as the primary means of measurement.

\section{Theoretical formulations}
We  consider a bipartite system consisting of two quantum dots A and B. For a bipartite states system ``AB'' a pure state can be written as a double sum over the product basis $\{|u_i\rangle\otimes|v_i\rangle\}$, as \cite{meystre2007elements, giovannetti2003characterizing}
\begin{equation}\label{eq:3}
|{\psi}\rangle_{AB}=\sum_{i,j}c_{i,j}|u_i\rangle_A\otimes|v_j\rangle_B.
\end{equation}
The QDs represent two qubits consisting of four electronics states, Viz; heavy hole valence band states ($|\pm\frac{3}{2}\rangle$ and conduction band states $|\pm\frac{1}{2}\rangle$. A composite system which has $4$-states can represent two qubits as \cite{diosi2011short}.
\begin{equation}\label{eq:4}
|{\psi}\rangle_{AB}=\sum_{\lambda}\sum_{l}c_{\lambda,l}|\lambda; A\rangle\otimes|l; B\rangle,
\end{equation}
where $\{\lambda; A\rangle\}$ and $\{l; B\rangle\}$ are certain angular momentum bases in the respective subsystems
A and B.

These states in quantum system can be alternatively represent by density operator. Use of density operator has significant advantage for relaxation and nonlinear quantum optics.
The density matrix or density operator defined as the outer product of the state
\begin{equation}\label{eq:5}
\hat{\rho}(t)={|\psi(t)\rangle}{\langle\psi(t)|}. 
\end{equation}
Thus a pure state contains all the information about the system using single state vector. The mixed state of the density matrix is defined as
 \begin{equation}\label{eq:6}
\hat{\rho}(t)=\sum_{k=1}^{n}{p_k}{|\psi(t)_k\rangle}{\langle\psi(t)_k|},
\end{equation}
where, $p_k$ is the probability distribution associated with each state vector $\psi_k$.
It is clearly seen from the definition of density matrix $\hat{\rho}$ of pure and mixed state that pure state is a special case of mixed state, where one of the probability ${p_k}$ has value equal to one and other have values equal to zero. This give us clear definition of incoherent mixture given by equation  ($\ref{eq:6}$).
The density matrix corresponding to coherent state $\sum_{k}{c_k}{|\psi_k\rangle}$ can be written as
 \begin{equation}\label{eq:7}
\hat{\rho}=\sum_{k}{|c_k|^2}{|\psi_k\rangle}{\langle\psi_k|}+\sum_{k{\neq}j}{c_k}^*{c_j}{|\psi_k\rangle}{\langle\psi_j|},
\end{equation}
One can notice that the second term in equation  ($\ref{eq:7}$) is interference term. 

The time evolution of density operator is govern by Schr\"odinger equation
 \begin{equation}\label{eq:8}
{i}{\hbar\ }\frac{\partial\hat{\rho}}{\partial{t}}=-[\rho, H],
\end{equation}
where $H$ is Hamiltonian of the interaction. In our case we considered the Hamiltonian for the case of  time evolution for the population in different energy states. which is defined as
\begin{equation}\label{eq:9}
H'= -\frac{1}{2}({\mu }^{\pm}E^{\mp}+ c.c.)  
\end{equation} 
where ${\mu }^{\pm }={\mu }_x\pm i{\mu }_y$ and $E^{\pm }=E_x\pm iE_y$.

We use the spin state of the DMS-QD, which can be split via strong magnetic field normal to growth axis in a QD. Here due to act of such consideration, the  
degeneracy of the $|\pm{\frac{3}{2}}\rangle$ as well as $|\pm{\frac{1}{2}}\rangle$ states are lifted and these four levels served as the 2 two-levels system. Population and electron spin mechanism manipulated through pulse sequences. The pulses considered here are sequentially operated, left and right circular polarized form of a laser, which excites electron from  ${J_z}=|\frac{3}{2}\rangle$ state in valence band
(VB) to generate population in the ${J_z}=|\frac{1}{2}\rangle$ state in conduction band
(CB). 

\begin{figure}[!htbp] 
    \centering
\includegraphics[width=.85\columnwidth]{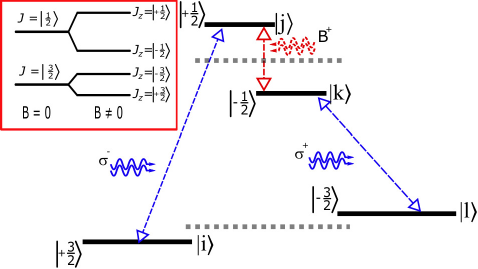}
\caption{Optical and magnetic assisted transitions processes considered are exhibited through a $\Lambda$-type four level system.}\label{Fig. 1}
\end{figure} 

As exhibited in Fig. \ref{Fig. 1}, we considered the interaction of three pulses in sequence with the quantum dot under study. First optical pulse raises the population from $|+\frac{3}{2}\rangle$ $(|ii\rangle)$ to $|+\frac{1}{2}\rangle$ $(|jj\rangle)$  via a right circularly polarized pulse (indicated as $\sigma^-$). The second pulse is magnetic pulse ($B^{+}$) at resonance energy of spin split energy of  conduction band (CB) states responsible for spin flip in QD, raises the population from $|+\frac{1}{2}\rangle$ $(|jj\rangle)$ to $|-\frac{1}{2}\rangle$ $(|kk\rangle)$ . The final third pulse raises the population of the electron from $|-\frac{1}{2}\rangle$ $(|kk\rangle)$  to $|-\frac{3}{2}\rangle$ $(|ll\rangle)$  via $\sigma^+$ which is left circularly polarized pulse. 

In practice, it is difficult to grow and illuminate a isolated single quantum dot. We considered that an ensemble of quantum dots which get illuminated due to finite spot size of the laser. The entanglement of states between two quantum dots will be obtained in the forthcoming discussions. To establish the entanglement between the energy spin states we define the wave-function for the two different bipartite system, for the first bipartite system we consider the energy states, $|\frac{3}{2}\rangle, |\frac{1}{2}\rangle, |-\frac{1}{2}\rangle$ and $|-\frac{3}{2}\rangle$ of first quantum dot  and similarly we consider for the another second quantum dot. 
\begin{figure}
    \centering
\includegraphics[width=.85\columnwidth]{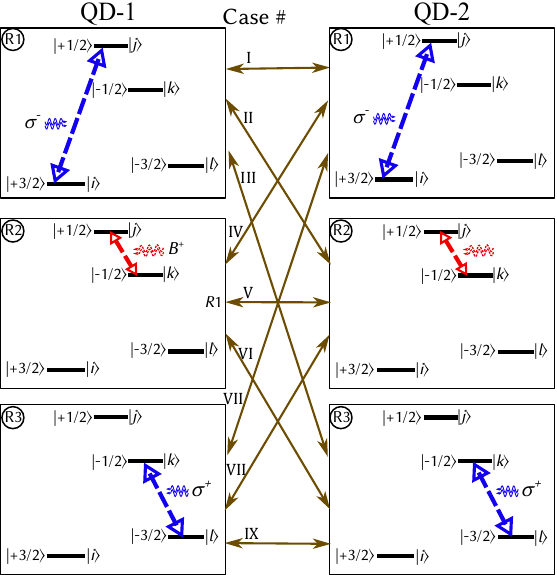}
\caption{Possible nine-bipartite systems (case I\ to IX) are exhibited. The quantum dots 1 and 2 may be prepared in any one of the subsystems R1, R2 or R3. Detailed population dynamics the subsystem are discussed in \cite{mehta2021room}.}\label{Fig. 2}
\end{figure} 

\subsection{Preparation of Quantum Dot}

As shown in Fig. \ref{Fig. 1}, the QD having $\Lambda$ type energy structure can be excited by an appropriate optical pulse of $\sigma^-$ polarization scuh that the 
population goes from state $|+\frac{3}{2}\rangle \leftrightarrows |+\frac{1}{2}\rangle$ states. 
 Also a magnetic pulse energy can create population oscillation between $|+\frac{1}{2}\rangle \leftrightarrows |-\frac{1}{2}\rangle$ states.
On the other hand, optical excitation of $\sigma^+$ polarization exhibits population oscillation between $|-\frac{3}{2}\rangle \leftrightarrows |-\frac{1}{2}\rangle$ states. 
The preparation of quantum dots in these subsystems are termed as R1, R2 and R3 respectively.

If two independent quantum dots prepared in any one of these subsystems are considered to be in a bipartite system. Together they form nine combinations as shown in Fig. \ref{Fig. 2}. 
 The sequential illumination of the quantum dots with optical and magnetic pulses may keep the QDs in the subsystem for times less than the relaxation time. Our objective is to study the entanglement between these energy states. 
\section{Results and Discussions}
In general the wavefunction for the bipartite system created from two subsystems $A$ and $B$ is defined as
\begin{equation}
|{\psi}\rangle= \sum{p_{A,B}|{\psi}{\rangle}_A{\otimes}|{\psi}{\rangle}_B}.\label{eq:14}
\end{equation}
here $p_{{A},{B}}$ is probability coefficient. Detailed derivations are given in \ref{appA}.

The density matrix operator $\hat{\rho }(t)$ corresponding to the matter-field interactions between the spin split levels responsible for the transition between two-level system. We use the master equation 
\begin{equation}
\hat{\dot{\rho}}(t)=-\frac{i}{\hbar\ }[H(t),\;\hat{\rho}(t)]-\Gamma\hat{\rho}(t),\label{rho}
\end{equation}
where $\Gamma$ is the phenomenological decay constant.
The  Hamiltonian $H(t) (=H_0 +H'(t))$ and density matrix $\hat\rho(t)$ for nine different bipartite systems are expressed as
\begin{equation}\label{eq:16}
\hat\rho(t)=|{\psi_{A}}\rangle\langle{\psi_{B}}|,
\end{equation}

The eigenvalues for the reduced density matrices are in the form of 
\begin{equation}\label{eq:31}
\lambda_{A,B}=\frac{1}{2}(1\pm\sqrt{(1-C^2)},
\end{equation}
here $C$ is concurrence. 
The entanglement (E) in pure state is measured by its entropy of entanglement,
\begin{equation}\label{eq:33}
E=S(\rho_A)=S(\rho_B),
\end{equation}
where $S(\rho)$ is the von Neumann entropy defined as $-Tr\rho{\log_2 \rho}$.
The entanglement can be written in the form of eigenvalues for the reduced density matrices as  
\begin{equation}\label{eq:34}
E=-\lambda_A\log_{2}\lambda_A-\lambda_B\log_{2}\lambda_B.
\end{equation}
Detailed calculations of concurrence and entanglement are given in \ref{appB}.

\begin{figure}[h]
    \centering
\includegraphics[width=1\columnwidth]{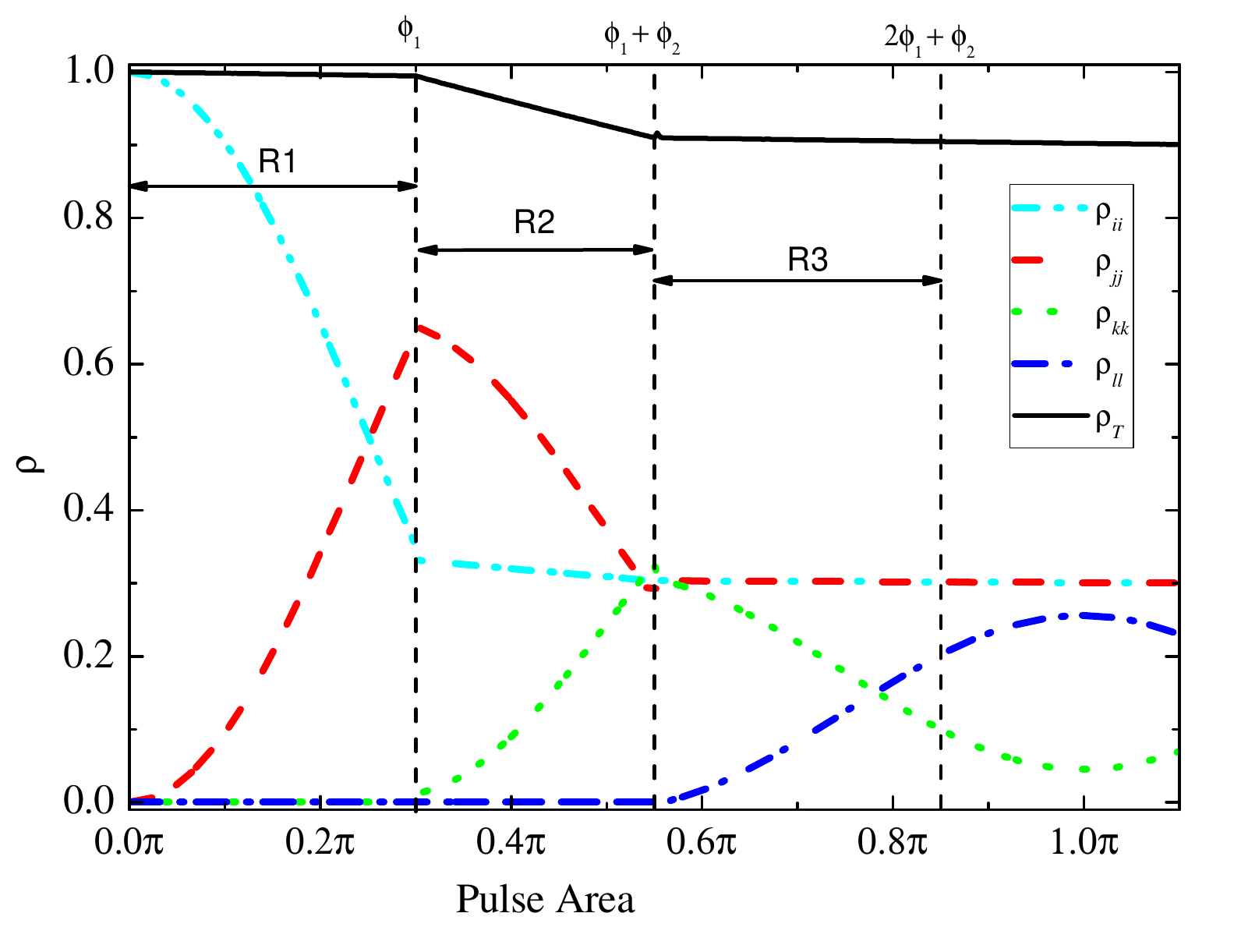}
\caption{Population dynamics of all four states in the $\Lambda$ system, considered. The total population $\rho_T$ represented as solid line shows a decay due to the relaxation process accounted in the calculations. We represented the state of DMS QD as `R1' during the shining of area pulse $\phi$, while the system is represented as `R2' during the area pulse of ($\phi_1+\phi_2$) and if the optical excitation pulse are of  ($2\phi_1+\phi_2$) leads to a subsystem R3.}\label{Fig. 3}
\end{figure}

\subsection{Population dynamics}
The population dynamics of the QDs in electronic states is shown in Fig. \ref{Fig. 3}.  
The population from the state $|+\frac{3}{2}\rangle$ (or $\rho_{ii}$) is excited to state $|+\frac{1}{2}\rangle$ (or $\rho_{jj}$) via right circular polarization. The evolution of population dynamics is addressed in detail in our previously published work \cite{mehta2021room}. However, in the present situation we are interested in understanding the population dynamics only up to $\pi$-pulse. We studied here the population dynamics for entanglement between the QDs, which is the requirement for microscopic study. The pulse duration for the first regime that last up-to ${\phi}_{1}=(\frac{\pi}{3})$ pulse area as  depicted in Fig. \ref{Fig. 3} shows that the population in $|+\frac{3}{2}\rangle$ state as $\rho_{ii}$ decreases, while the population in $|+\frac{1}{2}\rangle$ described by $\rho_{jj}$ increases. At the termination of $1$st pulse at $\frac{\pi}{3}$ pulse area. Second pulse begin at ${\phi}_{1}$ that last up-to ${\phi}_{1}+ {\phi}_{2}=(\frac{\pi}{3}+\frac{\pi}{4})$ the population from $|+\frac{1}{2}\rangle$ state raises to $|-\frac{1}{2}\rangle$ to state. This occurred because the second pulse, which is a magnetic pulse flips the spin state. One can notice from Fig. \ref{Fig. 3} that the population $\rho_{kk}$ in state $|-\frac{1}{2}\rangle$ exponentially increases during the excitation by second pulse. At the termination of $2$nd pulse at ${\phi}_{1}+ {\phi}_{2}$ the third pulse which is left circular polarize pulse deexcites the population from $\rho_{kk}$ to $\rho_{ll}$, up to the pulse area $2\frac{\pi}{3}+\frac{\pi}{4}={2}{\phi}_{1}+{\phi}_{2}$. Fig. \ref{Fig. 3} shows the true nature of population govern in such system. We have also plotted the total population  ${\rho}_{T}$  as a function of pulse area in the same Fig. \ref{Fig. 3} and it can be observed that the total population decreases due to incoherent noise.

\begin{figure}[h]
    \centering
\includegraphics[width=1\columnwidth]{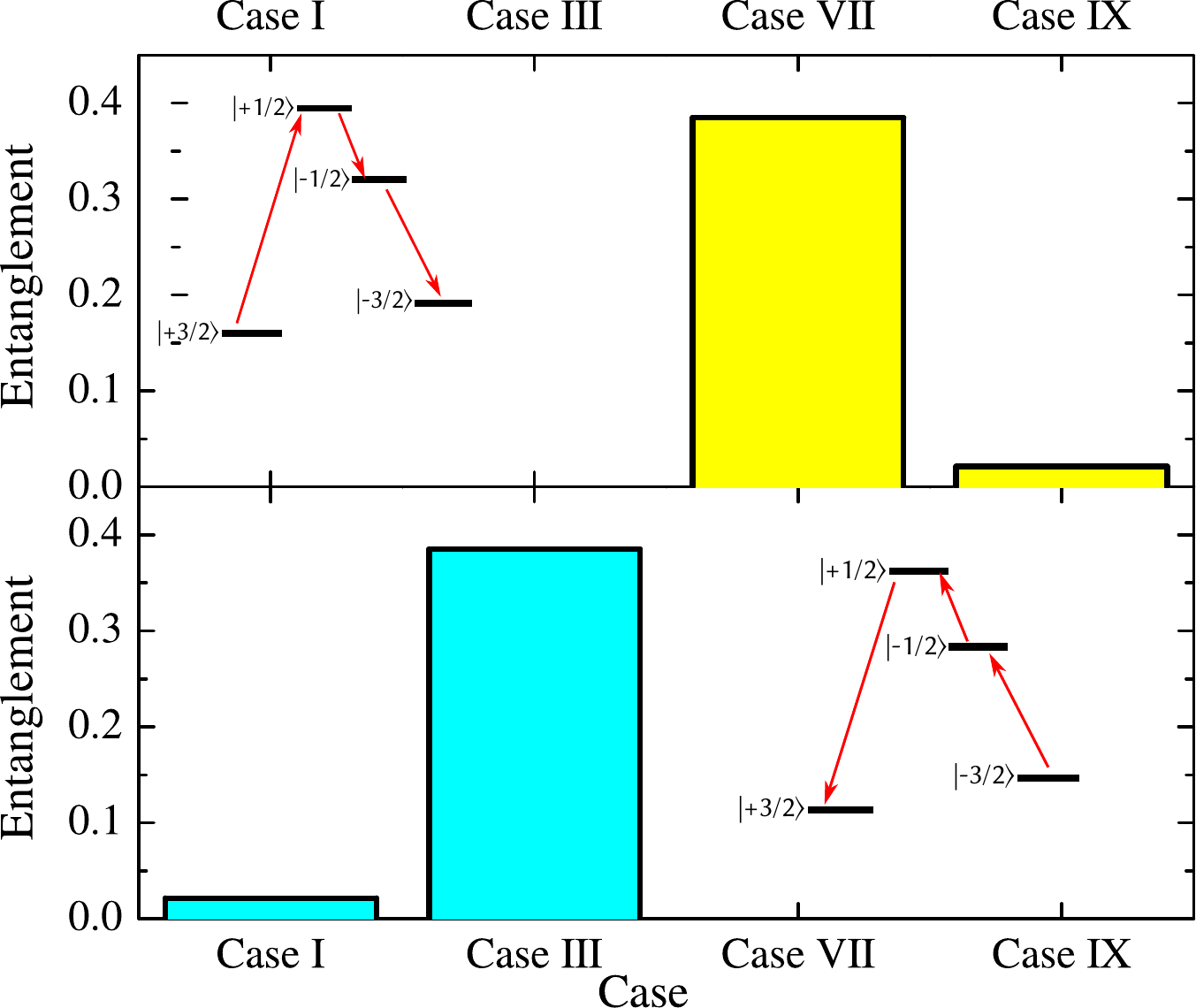}
\caption{Degree of entanglement for various bipartite states is exhibited. The QD in a bipartite state `A' and another QD in a bipartite state `B' as shown in \ref{Fig. 3} are entangled. Four possible entanglement are reported here. (i) Both QDs in bipartite state `A', represented as $A_1\rightarrow A_2$,  (ii) Both QDs in bipartite state `B', represented as $B_1\rightarrow B_2$, and (iii)/(iv) one QD in state `A' and the other in `B' represented by $A_1\rightarrow B_2$  and $B_1\rightarrow A_2$.}
\label{Fig.4}
\end{figure}

The parameters taken here to calculate the entanglement are as follows, the time duration taken for the excitation for the first pulse is ${\tau}_1=0.3$ ps and ${\tau}_2$ duration for the second pulse. This second pulse flip the spin from $|+\frac{1}{2}\rangle$ to $|-\frac{1}{2}\rangle$ for the time duration is $3.4$ ps. The third pulse duration for the de-excitation of population from $|-\frac{1}{2}\rangle$ to $|-\frac{3}{2}\rangle$ is $5.4$ ps which last up to the pulse area $1.1\pi$. The interaction Hamiltonian for radiation-matter defined as $H^{'}= \hbar\Omega_{ij}$. Here $\Omega_{ij}$ is Rabi flopping frequencies defined as $\Omega^{+}_{01}(t)=\frac{\mu_{01}^{+}E^{-}(t)}{\hbar}$ for the first pulse, $\Omega^{}_{12}(t)=\frac{\wp_{12}B(t)}{\hbar}$ for the second pulse and $\Omega^{-}_{23}(t)=\frac{\mu_{23}^{-}E^{+}(t)}{\hbar}$ is the Rabi flopping frequency for the third pulse. The transition dipole moment for electric and magnetic fields are $\mu_{ij}$ and $\wp_{ij}$ which is taken as phenomenological.

\subsection{Concurrence and Entanglement}
 Using the above parameters the entanglement for Case-I is obtained from equation (\ref{eq:35}) and it is found to be $6.13\times10^{-3}$, this value is not much significant, but the possibility of entanglement of the state can't be ruled out. We have then obtained the entanglement for the Case III, here the quantum dots are polarization entangled via spontaneous optical stimulation via right and left circularly polarized pulsed in first and second quantum dots. The value for the entanglement between the QDs is $0.385$ which comes out through equation (\ref{eq:43}). Further we moved to the Case VII. The quantum dots are entangled via left and right circularly polarized pulsed in the first and second quantum dots. The entanglement shows the remarkable significant value up-to the scale to attract the results is $0.021$, the output is the result obtained from equation (\ref{eq:59}). The last case we investigate between the QDs is the Case IX to observe the entanglement. Here we choose the QDs for which the electronic states for the QDs are $|-\frac{1}{2}\rangle$ and $|-\frac{3}{2}\rangle$. Both are governed by left circularly polarized light. The result of the entanglement as lead by the equation (\ref{eq:67}) is $6.159\times10^{-4}$. We plotted the entanglement of these two QDs in a different time of optical pulses in Fig. \ref{Fig.4}. We found that, in these four cases the Case III and Case VII show the entanglement of QDs with the executable results. These results also show the strong correlation between the QDs, when we reverse the execution of polarization pulses as depicted in the inset of fig. \ref{Fig.4}. The horizontal axis in figure \ref{Fig.4}, gives the details of  two cases where we have entanglement found between the quantum dots. The measurement of entanglement between two qubits can be realized by the concurrence ``C''. Due to finite value of concurrence, we got the entanglement in our cases. The Bell state corresponds to entangleThomasment in our case is $\frac{{|0\rangle}_{A}{|1\rangle}_{B}+{|1\rangle}_{B}{|0\rangle}_{A}}{\sqrt{2}}$.

\section{Conclusions}
We considered the bipartite system of two quantum dots simultaneously excited by the sequence of pulses. In the proposed preparation of the electronic states in QDs as discussed earlier, the population in the $|+\frac{1}{2}\rangle$ conduction band state is obtained via the optical pulse, the $|-\frac{1}{2}\rangle$ state population results due to the magnetic pulse which causes spin flipping of $|+\frac{1}{2}\rangle$ electron in the state. The simultaneous excitation of the quantum dots may lead to the nine possibilities of excitations as shown in Fig. \ref{Fig. 2}. The deexcitation of electrons in $|{\pm}\frac{3}{2}\rangle$ states will lead to the emission of  right and left circularly polarized photons. We have obtained the expressions for the entanglement as well as energy states of emitted photons resulting due to these deexcitations.

\section{Acknowledgement}
The authors thank late Dr. P. K. Sen for discussions. JTA acknowledge the financial support received from RPS-AICTE, New Delhi and SERB-DST,
New Delhi.

\appendix
\section{Density Matrix}\label{appA}
General wave function for the Case $I$ (see Fig. \ref{Fig. 2}), can be written as

\begin{equation}\label{eq:10}
|{\psi}\rangle_{1}=a|\frac{3}{2}\rangle+b|\frac{1}{2}\rangle,
\end{equation}
 and the wave function for the second QD is
\begin{equation}\label{eq:11}
|{\psi}\rangle_{2}=c|\frac{3}{2}\rangle+d|\frac{1}{2}\rangle.
\end{equation}
Here $a, b, c,$ and  $d$ are probability coefficients associated with each energy states.
The bipartite states together for the system can be written as
\begin{equation}\label{eq:12}
|{\psi}\rangle_{I}=|{\psi}\rangle_{1}\otimes|{\psi}\rangle_{2}.
\end{equation}
\begin{eqnarray}\label{eq:13}
|{\psi}\rangle &=&a^{'}|\frac{3}{2}\rangle\otimes|\frac{3}{2}\rangle+b^{'}|\frac{3}{2}\rangle\otimes|\frac{1}{2}\rangle \nonumber\\
 && +c^{'}|\frac{1}{2}\rangle\otimes|\frac{3}{2}\rangle+d^{'}|\frac{1}{2}\rangle\otimes|\frac{1}{2}\rangle. 
\end{eqnarray}

where $a^{'}, b^{'}, c^{'}$, and  $d^{'}$ are probability coefficients associated with bipartite energy states.

In general this can be written as,
\begin{equation}
|{\psi}\rangle=\qquad \sum{p_{\alpha,\beta}|{\psi}{\rangle}_\alpha{\otimes}|{\psi}{\rangle}_\beta}.\label{eq:14a}
\end{equation}
here $p_{{\alpha}{\beta}}$ is probability coefficient. 

The density matrix operator $\hat{\rho }(t)$ corresponding to the matter-radiation interactions between the spin split levels responsible for the each different bipartite system is given as,
\begin{eqnarray}
\hat\rho(t)_{I}= &\left(a{'}|\frac{3}{2}\rangle\otimes|\frac{3}{2}\rangle+b^{'}|\frac{3}{2}\rangle\otimes|\frac{1}{2}\rangle\right. \nonumber \\
& \left. +c^{'}|\frac{1}{2}\rangle\otimes|\frac{3}{2}\rangle+d^{'}|\frac{1}{2}\rangle\otimes|\frac{1}{2}\rangle\right) \nonumber\\
& \left(a'^{*}\langle\frac{3}{2}|\otimes\langle\frac{3}{2}|+b^{'*}\langle\frac{3}{2}|\otimes\langle\frac{1}{2}|\right.\nonumber \\
& \left. +c^{'*}\langle\frac{1}{2}|\otimes\langle\frac{3}{2}|+d^{'*}\langle\frac{1}{2}|\otimes\langle\frac{1}{2}|\right),
\label{eq:17}
\end{eqnarray}
\begin{equation}\label{eq:18}
\hat\rho(t)_{I}=\begin{pmatrix}
a^{'}a^{'*} & a^{'}b^{'*}&a^{'}c^{'*}&a^{'}d^{'*} \\                                                  
b^{'}a^{'*} & b^{'}b^{'*}&b^{'}c^{'*}&b^{'}d^{'*} \\      
c^{'}a^{'*} & c^{'}b^{'*}&c^{'}c^{'*}&c^{'}d^{'*} \\      
d^{'}a^{'*} & d^{'}b^{'*}&d^{'}c^{'*}&d^{'}d^{'*}                                           
\end{pmatrix},
\end{equation}
before we go further ahead in our approach to study the entanglement of the bipartite system. We generalized the energy levels to ease for writing the density matrix of the nine combinations. We shall consider the present spin state representation as $|\frac{3}{2}\rangle \Rightarrow |i\rangle$, $|\frac{1}{2}\rangle \Rightarrow |j\rangle$, $|-\frac{1}{2}\rangle \Rightarrow |k\rangle$ and  $|-\frac{3}{2}\rangle \Rightarrow |l\rangle$. 

So the density matrix for the first case can be written as
\begin{equation}\label{eq:19}
\hat\rho(t)_{I}=\begin{pmatrix}
\rho_{ii}\rho_{ii}^* & \rho_{ii}\rho_{ij}^*&\rho_{ij}\rho_{ii}^*&\rho_{ij}\rho_{ij}^* \\                                                 
\rho_{ii}\rho_{ji}^* &\rho_{ii}\rho_{jj}^*&\rho_{ij}\rho_{ji}^*&\rho_{ij}\rho_{jj}^* \\       
\rho_{ji}\rho_{ii}^* &\rho_{ji}\rho_{ij}^*&\rho_{jj}\rho_{ii}^*&\rho_{jj}\rho_{ij}^* \\       
\rho_{ji}\rho_{ji}^* &\rho_{ji}\rho_{jj}^*&\rho_{jj}\rho_{ji}^*&\rho_{jj}\rho_{jj}^*                                  
\end{pmatrix},
\end{equation}
correspondingly, above procedure can be imposed to write the density matrix for the rest of the cases, we get 
\begin{equation}\label{eq:20}
\hat\rho(t)_{II}=\begin{pmatrix}
\rho_{ii}\rho_{jj}^* & \rho_{ii}\rho_{jk}^*&\rho_{ij}\rho_{jj}^*&\rho_{ij}\rho_{jk}^* \\                                                 
\rho_{ii}\rho_{kj}^* &\rho_{ii}\rho_{kk}^*&\rho_{ij}\rho_{kj}^*&\rho_{ij}\rho_{kk}^* \\       
\rho_{ji}\rho_{jj}^* &\rho_{ji}\rho_{jk}^*&\rho_{jj}\rho_{jj}^*&\rho_{jj}\rho_{jk}^* \\       
\rho_{ji}\rho_{kj}^* &\rho_{ji}\rho_{kk}^*&\rho_{jj}\rho_{kj}^*&\rho_{jj}\rho_{kk}^*                                  
\end{pmatrix},
\end{equation}\break
\begin{equation}\label{eq:21}
    \hat\rho(t)_{III}=\begin{pmatrix}
\rho_{ii}\rho_{kk}^* & \rho_{ii}\rho_{kl}^*&\rho_{ij}\rho_{kk}^*&\rho_{ij}\rho_{kl}^* \\                                                 
\rho_{ii}\rho_{lk}^* &\rho_{ii}\rho_{ll}^*&\rho_{ij}\rho_{lk}^*&\rho_{ij}\rho_{ll}^* \\       
\rho_{ji}\rho_{kk}^* &\rho_{ji}\rho_{kl}^*&\rho_{jj}\rho_{kk}^*&\rho_{jj}\rho_{kl}^* \\       
\rho_{ji}\rho_{lk}^* &\rho_{ji}\rho_{ll}^*&\rho_{jj}\rho_{lk}^*&\rho_{jj}\rho_{ll}^*                                  
\end{pmatrix},
  \end{equation}
\begin{equation}\label{eq:22}
\hat\rho(t)_{IV}=\begin{pmatrix}
\rho_{jj}\rho_{ii}^* & \rho_{jj}\rho_{ij}^*&\rho_{jk}\rho_{ii}^*&\rho_{jk}\rho_{ij}^* \\                                                 
\rho_{jj}\rho_{ji}^* &\rho_{jj}\rho_{jj}^*&\rho_{jk}\rho_{ji}^*&\rho_{jk}\rho_{jj}^* \\       
\rho_{kj}\rho_{ii}^* &\rho_{kj}\rho_{ij}^*&\rho_{kk}\rho_{ii}^*&\rho_{kk}\rho_{ij}^* \\       
\rho_{kj}\rho_{ji}^* &\rho_{kj}\rho_{jj}^*&\rho_{kk}\rho_{ji}^*&\rho_{kk}\rho_{jj}^*                                  
\end{pmatrix},
\end{equation}\break
\begin{equation}\label{eq:23}
    \hat\rho(t)_{V}=\begin{pmatrix}
\rho_{jj}\rho_{jj}^* & \rho_{jj}\rho_{jk}^*&\rho_{jk}\rho_{jj}^*&\rho_{jk}\rho_{jk}^* \\                                                 
\rho_{jj}\rho_{kj}^* &\rho_{jj}\rho_{kk}^*&\rho_{jk}\rho_{kj}^*&\rho_{jk}\rho_{kk}^* \\       
\rho_{kj}\rho_{jj}^* &\rho_{kj}\rho_{jk}^*&\rho_{kk}\rho_{jj}^*&\rho_{kk}\rho_{jk}^* \\       
\rho_{kj}\rho_{kj}^* &\rho_{kj}\rho_{kk}^*&\rho_{kk}\rho_{kj}^*&\rho_{kk}\rho_{kk}^*                                  
\end{pmatrix},
\end{equation}
\begin{equation}\label{eq:24}
\hat\rho(t)_{VI}=\begin{pmatrix}
\rho_{jj}\rho_{kk}^* & \rho_{jj}\rho_{kl}^*&\rho_{jk}\rho_{kk}^*&\rho_{jk}\rho_{kl}^* \\                                                 
\rho_{jj}\rho_{lk}^* &\rho_{jj}\rho_{ll}^*&\rho_{jk}\rho_{lk}^*&\rho_{jk}\rho_{ll}^* \\       
\rho_{kj}\rho_{kk}^* &\rho_{kj}\rho_{kl}^*&\rho_{kk}\rho_{kk}^*&\rho_{kk}\rho_{kl}^* \\       
\rho_{kj}\rho_{lk}^* &\rho_{kj}\rho_{ll}^*&\rho_{kk}\rho_{lk}^*&\rho_{kk}\rho_{ll}^*                                  
\end{pmatrix},
\end{equation}
\begin{equation}\label{eq:25}
    \hat\rho(t)_{VII}=\begin{pmatrix}
\rho_{kk}\rho_{ii}^* & \rho_{kk}\rho_{ij}^*&\rho_{kl}\rho_{ii}^*&\rho_{kl}\rho_{ij}^* \\                                                 
\rho_{kk}\rho_{ji}^* &\rho_{kk}\rho_{jj}^*&\rho_{kl}\rho_{ji}^*&\rho_{kl}\rho_{jj}^* \\       
\rho_{lk}\rho_{ii}^* &\rho_{lk}\rho_{ij}^*&\rho_{ll}\rho_{ii}^*&\rho_{ll}\rho_{ij}^* \\       
\rho_{lk}\rho_{ji}^* &\rho_{lk}\rho_{jj}^*&\rho_{ll}\rho_{ji}^*&\rho_{ll}\rho_{jj}^*                                  
\end{pmatrix},
  \end{equation}
\begin{equation}\label{eq:26}
\hat\rho(t)_{VIII}=\begin{pmatrix}
\rho_{kk}\rho_{jj}^* & \rho_{kk}\rho_{jk}^*&\rho_{kl}\rho_{jj}^*&\rho_{kl}\rho_{jk}^* \\                                                 
\rho_{kk}\rho_{kj}^* &\rho_{kk}\rho_{kk}^*&\rho_{kl}\rho_{kj}^*&\rho_{kl}\rho_{jk}^* \\       
\rho_{lk}\rho_{jj}^* &\rho_{lk}\rho_{jk}^*&\rho_{ll}\rho_{jj}^*&\rho_{ll}\rho_{jk}^* \\       
\rho_{lk}\rho_{kj}^* &\rho_{lk}\rho_{kk}^*&\rho_{ll}\rho_{kj}^*&\rho_{ll}\rho_{kk}^*                                  
\end{pmatrix},
\end{equation}
\begin{equation}\label{eq:27}
    \hat\rho(t)_{IX}=\begin{pmatrix}
\rho_{kk}\rho_{kk}^* & \rho_{kk}\rho_{kl}^*&\rho_{kl}\rho_{kk}^*&\rho_{kl}\rho_{kl}^* \\                                                 
\rho_{kk}\rho_{lk}^* &\rho_{kk}\rho_{ll}^*&\rho_{kl}\rho_{lk}^*&\rho_{kl}\rho_{ll}^* \\       
\rho_{lk}\rho_{kk}^* &\rho_{lk}\rho_{kl}^*&\rho_{ll}\rho_{kk}^*&\rho_{ll}\rho_{kl}^* \\       
\rho_{lk}\rho_{lk}^* &\rho_{lk}\rho_{ll}^*&\rho_{ll}\rho_{lk}^*&\rho_{ll}\rho_{ll}^*                                  
\end{pmatrix}.
  \end{equation}

The wave function of the qubit system in equation \ref{eq:13}, can be written in $|i\rangle, |j\rangle, |k\rangle$ and $|l\rangle$ generalized basis, 
\begin{equation}\label{eq:28}
{|{\psi}\rangle_{}=a^{'}|i\rangle\otimes|i\rangle+b^{'}i\rangle\otimes|j\rangle+c^{'}|j\rangle\otimes|i\rangle+d^{'}|j\rangle\otimes|j\rangle,} 
\end{equation}
where $|a'|^2+|b'|^2+|c'|^2+|d'|^2=1$, that implies for the density matrix of the desired case.
In view of this we can write,
one of the eigenvalue is $1$ for the eigenvector of the matrix in hand and the other three eigenvalues are zero \cite{aldoshin2014quantum}. In order to see the effect of entanglement in energy states, we need to write the reduced density matrices for each and every case, before we start further investigation. The reduced density matrices for the case $I$ are

\begin{equation}\label{eq:29}
\hat\rho(t)_{AI}=\begin{pmatrix}
\rho_{ii}\rho_{ii}^*+\rho_{ii}\rho_{jj}^*&\rho_{ij}\rho_{ii}^*+\rho_{ij}\rho_{jj}^*\\
\rho_{ji}\rho_{ii}^*+\rho_{ji}\rho_{jj}^*& \rho_{jj}\rho_{ii}^*+\rho_{jj}\rho_{jj}^*                                
\end{pmatrix},
\end{equation}

\begin{equation}\label{eq:30}
\hat\rho(t)_{BI}=\begin{pmatrix}
\rho_{ii}\rho_{ii}^*+\rho_{jj}\rho_{ii}^*  &\rho_{ii}\rho_{ij}^*+\rho_{jj}\rho_{ij}^* \\
\rho_{ii}\rho_{ji}^*+\rho_{jj}\rho_{ji}^*& \rho_{ii}\rho_{jj}^*+\rho_{jj}\rho_{jj}^*                                
\end{pmatrix}.
\end{equation}

\section{Concurrence}\label{appB}
The entanglement for the first case is 
\begin{equation}\label{eq:35}
E_{I}=-\lambda_{AI}\log_{2}\left(\lambda_{AI}\right)-\lambda_{BI}\log_{2}\left(\lambda_{BI}\right).
\end{equation}
Same procedure implies to second and rest of the cases. We get the reduced density matrices for the subsequence cases as,
\begin{equation}\label{eq:36}
\hat\rho(t)_{AII}=\begin{pmatrix}
\rho_{ii}\rho_{jj}^*+\rho_{ii}\rho_{kk}^*&\rho_{ij}\rho_{jj}^*+\rho_{ij}\rho_{kk}^*\\
\rho_{ji}\rho_{jj}^*+\rho_{ji}\rho_{kk}^*& \rho_{jj}\rho_{jj}^*+\rho_{jj}\rho_{kk}^*                                
\end{pmatrix},
\end{equation}

\begin{equation}\label{eq:37}
\hat\rho(t)_{BII}=\begin{pmatrix}
\rho_{ii}\rho_{jj}^*+\rho_{jj}\rho_{jj}^*  &\rho_{ii}\rho_{jk}^*+\rho_{jj}\rho_{jk}^* \\
\rho_{ii}\rho_{kj}^*+\rho_{jj}\rho_{kj}^*& \rho_{ii}\rho_{kk}^*+\rho_{jj}\rho_{kk}^*                                
\end{pmatrix},
\end{equation}
and the concurrence $C_{II}$ is
\begin{eqnarray}\label{eq:38}
C_{II}&=&[(\rho_{ii}\rho_{jj}^*+\rho_{ii}\rho_{kk}^*)(\rho_{jj}\rho_{jj}^*+\rho_{jj}\rho_{kk}^*) \nonumber\\
&&\hspace{-5mm} - (\rho_{ij}\rho_{jj}^*+\rho_{ij}\rho_{kk}^*)(\rho_{ji}\rho_{jj}^*+\rho_{ji}\rho_{kk}^*)]^{1/2}.
\end{eqnarray}
The entanglement is 
\begin{equation}\label{eq:39}
E_{II}=-\lambda_{AII}\log_{2}\left(\lambda_{AII}\right)-\lambda_{BII}\log_{2}\left(\lambda_{BII}\right)
\end{equation}

\begin{equation}\label{eq:40}
\hat\rho(t)_{AIII}=\begin{pmatrix}
\rho_{ii}\rho_{kk}^*+\rho_{ii}\rho_{ll}^*&\rho_{ij}\rho_{kk}^*+\rho_{ij}\rho_{ll}^*\\
\rho_{ji}\rho_{kk}^*+\rho_{ji}\rho_{ll}^*& \rho_{jj}\rho_{kk}^*+\rho_{jj}\rho_{ll}^*                                
\end{pmatrix},
\end{equation}

\begin{equation}\label{eq:41}
\hat\rho(t)_{BIII}=\begin{pmatrix}
\rho_{ii}\rho_{kk}^*+\rho_{jj}\rho_{kk}^*  &\rho_{ii}\rho_{kl}^*+\rho_{jj}\rho_{kl}^* \\
\rho_{ii}\rho_{lk}^*+\rho_{jj}\rho_{lk}^*& \rho_{ii}\rho_{ll}^*+\rho_{jj}\rho_{ll}^*                                
\end{pmatrix},
\end{equation}
and the concurrence $C_{III}$ is
\begin{eqnarray}\label{eq:42}
C_{III}&=&[(\rho_{ii}\rho_{kk}^*+\rho_{ii}\rho_{ll}^*)(\rho_{jj}\rho_{kk}^*+\rho_{jj}\rho_{ll}^*) \nonumber\\
&&\hspace{-5mm}  (\rho_{ij}\rho_{kk}^*+\rho_{ij}\rho_{ll}^*)(\rho_{ji}\rho_{kk}^*+\rho_{ji}\rho_{ll}^*)]^{1/2}.
\end{eqnarray}
The entanglement is 
\begin{equation}\label{eq:43}
E_{III}=-\lambda_{AIII}\log_{2}\left(\lambda_{AIII}\right)-\lambda_{BIII}\log_{2}\left(\lambda_{BI}\right)
\end{equation}
For the case IV
\begin{equation}\label{eq:44}
\hat\rho(t)_{AIV}=\begin{pmatrix}
\rho_{jj}\rho_{ii}^*+\rho_{jj}\rho_{jj}^*&\rho_{jk}\rho_{ii}^*+\rho_{jk}\rho_{jj}^*\\
\rho_{kj}\rho_{ii}^*+\rho_{kj}\rho_{jj}^*& \rho_{kk}\rho_{ii}^*+\rho_{kk}\rho_{jj}^*                                
\end{pmatrix},
\end{equation}

\begin{equation}\label{eq:45}
\hat\rho(t)_{BIV}=\begin{pmatrix}
\rho_{jj}\rho_{ii}^*+\rho_{kk}\rho_{ii}^*  &\rho_{jj}\rho_{ij}^*+\rho_{kk}\rho_{ij}^* \\
\rho_{jj}\rho_{ji}^*+\rho_{kk}\rho_{ji}^*& \rho_{jj}\rho_{jj}^*+\rho_{kk}\rho_{jj}^*                                
\end{pmatrix},
\end{equation}
and the concurrence $C_{IV}$ is
\begin{eqnarray}\label{eq:46}
C_{IV}&=&[(\rho_{jj}\rho_{ii}^*+\rho_{jj}\rho_{jj}^*)(\rho_{kk}\rho_{ii}^*+\rho_{kk}\rho_{jj}^*) \nonumber\\
&&\hspace{-5mm} -(\rho_{jk}\rho_{ii}^*+\rho_{jk}\rho_{jj}^*)(\rho_{kj}\rho_{ii}^*+\rho_{kj}\rho_{jj}^*)]^{1/2}.
\end{eqnarray}
The entanglement is 
\begin{equation}\label{eq:47}
E_{IV}=-\lambda_{AIV}\log_{2}\left(\lambda_{AIV}\right)-\lambda_{BIV}\log_{2}\left(\lambda_{BIV}\right)
\end{equation}
For the case V
\begin{equation}\label{eq:48}
\hat\rho(t)_{AV}=\begin{pmatrix}
\rho_{jj}\rho_{jj}^*+\rho_{jj}\rho_{kk}^*&\rho_{jk}\rho_{jj}^*+\rho_{jk}\rho_{kk}^*\\
\rho_{kj}\rho_{jj}^*+\rho_{kj}\rho_{kk}^*& \rho_{kk}\rho_{jj}^*+\rho_{kk}\rho_{kk}^*                                
\end{pmatrix},
\end{equation}

\begin{equation}\label{eq:49}
\hat\rho(t)_{BV}=\begin{pmatrix}
\rho_{jj}\rho_{jj}^*+\rho_{kk}\rho_{jj}^*  &\rho_{jj}\rho_{jk}^*+\rho_{kk}\rho_{jk}^* \\
\rho_{jj}\rho_{kj}^*+\rho_{kk}\rho_{kj}^*& \rho_{jj}\rho_{kk}^*+\rho_{kk}\rho_{kk}^*                                
\end{pmatrix},
\end{equation}
and the concurrence $C_{V}$ is
\begin{eqnarray}\label{eq:50}
C_{V}&=&[(\rho_{jj}\rho_{jj}^*+\rho_{jj}\rho_{kk}^*)(\rho_{kk}\rho_{jj}^*+\rho_{kk}\rho_{kk}^*) \nonumber\\
&&\hspace{-6mm} - (\rho_{jk}\rho_{jj}^*+\rho_{jk}\rho_{kk}^*)(\rho_{kj}\rho_{jj}^*+\rho_{kj}\rho_{kk}^*)]^{1/2}.
\end{eqnarray}
The entanglement is 
\begin{equation}\label{eq:51}
E_{V}=-\lambda_{AV}\log_{2}\left(\lambda_{AV}\right)-\lambda_{BV}\log_{2}\left(\lambda_{BV}\right)
\end{equation}
For the case VI
\begin{equation}\label{eq:52}
\hat\rho(t)_{AVI}=\begin{pmatrix}
\rho_{jj}\rho_{kk}^*+\rho_{jj}\rho_{ll}^*&\rho_{jk}\rho_{kk}^*+\rho_{jk}\rho_{ll}^*\\
\rho_{kj}\rho_{kk}^*+\rho_{kj}\rho_{ll}^*& \rho_{kk}\rho_{kk}^*+\rho_{kk}\rho_{ll}^*                                
\end{pmatrix},
\end{equation}

\begin{equation}\label{eq:53}
\hat\rho(t)_{BVI}=\begin{pmatrix}
\rho_{jj}\rho_{kk}^*+\rho_{kk}\rho_{kk}^*  &\rho_{jj}\rho_{kl}^*+\rho_{kk}\rho_{kl}^* \\
\rho_{jj}\rho_{lk}^*+\rho_{kk}\rho_{lk}^*& \rho_{jj}\rho_{ll}^*+\rho_{kk}\rho_{ll}^*                                
\end{pmatrix},
\end{equation}
and the concurrence $C_{VI}$ is
\begin{eqnarray}\label{eq:54}
C_{VI}&=&[(\rho_{jj}\rho_{kk}^*+\rho_{jj}\rho_{ll}^*)(\rho_{kk}\rho_{kk}^*+\rho_{kk}\rho_{ll}^*) \nonumber\\
&&\hspace{-7mm} - (\rho_{jk}\rho_{kk}^*+\rho_{jk}\rho_{ll}^*)(\rho_{kj}\rho_{kk}^*+\rho_{kj}\rho_{ll}^*)]^{1/2}.
\end{eqnarray}
The entanglement is 
\begin{equation}\label{eq:55}
E_{VI}=-\lambda_{AVI}\log_{2}\left(\lambda_{AVI}\right)-\lambda_{BVI}\log_{2}\left(\lambda_{BVI}\right)
\end{equation}

For the case VII
\begin{equation}\label{eq:56}
\hat\rho(t)_{AVII}=\begin{pmatrix}
\rho_{kk}\rho_{ii}^*+\rho_{kk}\rho_{jj}^*&\rho_{kl}\rho_{ii}^*+\rho_{kl}\rho_{jj}^*\\
\rho_{lk}\rho_{ii}^*+\rho_{lk}\rho_{jj}^*& \rho_{ll}\rho_{ii}^*+\rho_{ll}\rho_{jj}^*                                
\end{pmatrix},
\end{equation}

\begin{equation}\label{eq:57}
\hat\rho(t)_{BVII}=\begin{pmatrix}
\rho_{kk}\rho_{ii}^*+\rho_{ll}\rho_{ii}^*  &\rho_{kk}\rho_{ij}^*+\rho_{ll}\rho_{ij}^* \\
\rho_{kk}\rho_{ji}^*+\rho_{ll}\rho_{ji}^*& \rho_{kk}\rho_{jj}^*+\rho_{ll}\rho_{jj}^*                                
\end{pmatrix},
\end{equation}
and the concurrence $C_{VII}$ is
\begin{eqnarray}\label{eq:58}
C_{VII}&=&[(\rho_{kk}\rho_{ii}^*+\rho_{kk}\rho_{jj}^*)(\rho_{ll}\rho_{ii}^*+\rho_{ll}\rho_{jj}^*) \nonumber\\
&&\hspace{-8mm} - (\rho_{kl}\rho_{ii}^*+\rho_{kl}\rho_{jj}^*)(\rho_{lk}\rho_{ii}^*+\rho_{lk}\rho_{jj}^*)]^{1/2}.
\end{eqnarray}
The entanglement is 
\begin{equation}\label{eq:59}
E_{VII}=-\lambda_{AVII}\log_{2}\left(\lambda_{AVII}\right)-\lambda_{BVII}\log_{2}\left(\lambda_{BVII}\right)
\end{equation}
For the case VIII
\begin{equation}\label{eq:60}
\hat\rho(t)_{AVIII}=\begin{pmatrix}
\rho_{kk}\rho_{jj}^*+\rho_{kk}\rho_{kk}^*&\rho_{kl}\rho_{jj}^*+\rho_{kl}\rho_{jk}^*\\
\rho_{lk}\rho_{jj}^*+\rho_{lk}\rho_{kk}^*& \rho_{ll}\rho_{jj}^*+\rho_{ll}\rho_{kk}^*                                
\end{pmatrix},
\end{equation}

\begin{equation}\label{eq:61}
\hat\rho(t)_{BVIII}=\begin{pmatrix}
\rho_{kk}\rho_{jj}^*+\rho_{ll}\rho_{jj}^*  &\rho_{kk}\rho_{jk}^*+\rho_{ll}\rho_{jk}^* \\
\rho_{kk}\rho_{kj}^*+\rho_{ll}\rho_{kj}^*& \rho_{kk}\rho_{kk}^*+\rho_{ll}\rho_{kk}^*                                
\end{pmatrix},
\end{equation}
and the concurrence $C_{VIII}$ is
\begin{eqnarray}\label{eq:62}
C_{VIII}&=&[(\rho_{kk}\rho_{jj}^*+\rho_{kk}\rho_{kk}^*)(\rho_{ll}\rho_{jj}^*+\rho_{ll}\rho_{kk}^*) \nonumber\\
&&\hspace{-8mm} - (\rho_{kl}\rho_{jj}^*+\rho_{kl}\rho_{jk}^*)(\rho_{lk}\rho_{jj}^*+\rho_{lk}\rho_{kk}^*)]^{1/2}.
\end{eqnarray}
The entanglement is 
\begin{equation}\label{eq:63}
E_{VIII}=-\lambda_{AVIII}\log_{2}\left(\lambda_{AVIII}\right)-\lambda_{BVIII}\log_{2}\left(\lambda_{BVIII}\right)
\end{equation}

For the case IX
\begin{equation}\label{eq:64}
\hat\rho(t)_{AIX}=\begin{pmatrix}
\rho_{kk}\rho_{kk}^*+\rho_{kk}\rho_{ll}^*&\rho_{kl}\rho_{kk}^*+\rho_{kl}\rho_{ll}^*\\
\rho_{lk}\rho_{kk}^*+\rho_{lk}\rho_{ll}^*& \rho_{ll}\rho_{kk}^*+\rho_{ll}\rho_{ll}^*                                
\end{pmatrix},
\end{equation}

\begin{equation}\label{eq:65}
\hat\rho(t)_{BIX}=\begin{pmatrix}
\rho_{kk}\rho_{kk}^*+\rho_{ll}\rho_{kk}^*  &\rho_{kk}\rho_{kl}^*+\rho_{kl}\rho_{ll}^* \\
\rho_{kk}\rho_{lk}^*+\rho_{ll}\rho_{lk}^*& \rho_{kk}\rho_{ll}^*+\rho_{ll}\rho_{ll}^*                                
\end{pmatrix},
\end{equation}
and the concurrence $C_{IX}$ is
\begin{eqnarray}\label{eq:66}
C_{IX}&=&[(\rho_{kk}\rho_{kk}^*+\rho_{kk}\rho_{ll}^*)( \rho_{ll}\rho_{kk}^*+\rho_{ll}\rho_{ll}^*) \nonumber\\
&&\hspace{-8mm} - (\rho_{kl}\rho_{kk}^*+\rho_{kl}\rho_{ll}^*)(\rho_{lk}\rho_{kk}^*+\rho_{lk}\rho_{ll}^*)]^{1/2}.
\end{eqnarray}
The entanglement is 
\begin{equation}\label{eq:67}
E_{IX}=-\lambda_{AIX}\log_{2}\left(\lambda_{AIX}\right)-\lambda_{BIX}\log_{2}\left(\lambda_{BIX}\right)
\end{equation}


\begin{thebibliography}{00}
\bibitem{einstein1935can} Einstein, A., Podolsky, B., \& Rosen, N. (1935). Can quantum-mechanical description of physical reality be considered complete?. Physical review, {47}(10), 777.
\bibitem{bennett1984quantum} Bennett, C. H., \& Brassard, G. (1984). Quantum Cryptography: Public Key Distribution and Coin Tossing. In Proceedings of IEEE International Conference on Computers, Systems and Signal Processing (pp. 175-179).
\bibitem{lewis2019dynamics} Lewis-Swan, R. J., Safavi-Naini, A., Kaufman, A. M., \& Rey, A. M. (2019). Dynamics of quantum information. Nature Reviews Physics, 1(10), 627-634.
\bibitem{hu2023progress} Hu, X. M., Guo, Y., Liu, B. H., Li, C. F., \& Guo, G. C. (2023). Progress in quantum teleportation. Nature Reviews Physics, 1-15.
\bibitem{bouwmeester2000physics} Bouwmeester, D., \& Zeilinger, A. (2000). The physics of quantum information: basic concepts. In The physics of quantum information: quantum cryptography, quantum teleportation, quantum computation (pp. 1-14). Berlin, Heidelberg: Springer Berlin Heidelberg.
\bibitem{bell1964einstein} Bell, J. S. (1964). On the einstein podolsky rosen paradox. Physics Physique Fizika, 1(3), 195.
\bibitem{benjamin2009prospects} Benjamin, S. C., Lovett, B. W., \& Smith, J. M. (2009). Prospects for measurement‐based quantum computing with solid state spins. Laser \& Photonics Reviews, 3(6), 556-574.
\bibitem{bogdanov2011quantum} Bogdanov, Y. I., Valiev, K. A., \& Kokin, A. A. (2011). Quantum computers: Achievements, implementation difficulties, and prospects. Russian Microelectronics, 40, 225-236.
\bibitem{nielsen2010quantum} Nielsen, M. A.,\& Chuang, I. L. (2010). Quantum computation and quantum information. Cambridge university press.
\bibitem{hughes2021entanglement} Hughes, C., Isaacson, J., Perry, A., Sun, R. F., Turner, J., Hughes, C., ... \& Turner, J. (2021). Entanglement (pp. 59-71). Springer International Publishing.
\bibitem{hu2021long} Hu, X. M., Huang, C. X., Sheng, Y. B., Zhou, L., Liu, B. H., Guo, Y., ... \& Guo, G. C. (2021). Long-distance entanglement purification for quantum communication. Physical review letters, 126(1), 010503.
\bibitem{kocher1967polarization} Kocher, C. A., \& Commins, E. D. (1967). Polarization correlation of photons emitted in an atomic cascade. Physical Review Letters, 18(15), 575.
\bibitem{togan2010nature} Togan, E., Chu, Y., Trifonov, A. S., Jiang, L., Maze, J., Childress, L., ... \& Lukin, M. D. (2010). Nature (London). Nature (London), 466(730).
\bibitem{warren2021robust} Warren, A., Güngördü, U., Kestner, J. P., Barnes, E., \& Economou, S. E. (2021). Robust photon-mediated entangling gates between quantum dot spin qubits. Physical Review B, 104(11), 115308.
\bibitem{chan2023chip} Chan, M. L., Tiranov, A., Appel, M. H., Wang, Y., Midolo, L., Scholz, S., ... \& Lodahl, P. (2023). On-chip spin-photon entanglement based on photon-scattering of a quantum dot. npj Quantum Information, 9(1), 49.
\bibitem{gupta2015entanglement} Gupta, V. P., Mandayam, P., Sunder, V. S., Gupta, V. P., Mandayam, P., \& Sunder, V. S. (2015). Entanglement in Bipartite Quantum States. The Functional Analysis of Quantum Information Theory: A Collection of Notes Based on Lectures by Gilles Pisier, KR Parthasarathy, Vern Paulsen and Andreas Winter, 39-62. 
\bibitem{roncaglia2014bipartite} Roncaglia, M., Montorsi, A., \& Genovese, M. (2014). Bipartite entanglement of quantum states in a pair basis. Physical Review A, 90(6), 062303.
\bibitem{meystre2007elements} Meystre, P., \& Sargent, M. (2007). Elements of quantum optics. Springer Science \& Business Media.
\bibitem{giovannetti2003characterizing} Giovannetti, V., Mancini, S., Vitali, D., and Tombesi, P. (2003). Characterizing the entanglement of bipartite quantum systems. Physical Review A, {67}(2), 022320.
\bibitem{aldoshin2014quantum} Aldoshin, S. M., Fel'dman, E. B., \& Yurishchev, M. A. (2014). Quantum entanglement and quantum discord in magnetoactive materials. Low Temperature Physics, {40}(1), 3-16.
\bibitem{diosi2011short} Diosi, L. (2011). A short course in quantum information theory: an approach from theoretical physics (Vol. 827). Springer.
\bibitem{mehta2021room} Mehta, M. K., Andrews, J. T., \& Sen, P. (2021). Room temperature CNOT operation in a diluted magnetic semiconductor quantum dot. Quantum Information Processing, 20(5), 194.
\end{thebibliography}
\end{document}